\documentclass[aps,prb,twocolumn,showpacs,superscriptaddress]{revtex4}

\usepackage{graphicx} 

\begin{document}

\title
{Reversible Band Gap Engineering in Carbon Nanotubes
by Radial Deformation}

\author{O. G\"{u}lseren}
\affiliation{NIST Center for Neutron Research,
National Institute of Standards and Technology,
Gaithersburg, MD 20899}
\affiliation{Department of Materials Science and Engineering,
University of Pennsylvania, Philadelphia, PA 19104}
\author{T. Yildirim}
\affiliation{NIST Center for Neutron Research,
National Institute of Standards and Technology,
Gaithersburg, MD 20899}
\author{S. Ciraci}
\affiliation{Department of Physics, Bilkent University,
Ankara 06533, Turkey}
\author{\c{C}. K{\i}l{\i}\c{c}}
\altaffiliation[Present address: ]{
National Renewable Energy Laboratory, Golden, CO 80401}
\affiliation{Department of Physics, Bilkent University,
Ankara 06533, Turkey}

\date{\today}

\begin{abstract}

We present a systematic analysis of the effect of radial deformation
on the atomic and electronic structure of zigzag and armchair single
wall carbon nanotubes using the first principle plane wave method.
The nanotubes were deformed by applying a radial strain, which distorts
the circular cross section to an elliptical one. The atomic structure of
the nanotubes under this strain are fully optimized, and the electronic
structure is calculated self-consistently to determine the response
of individual bands to the radial deformation. The band gap of the
insulating tube is closed and eventually an insulator-metal transition
sets in by the radial strain which is in the elastic range. Using this
property a multiple quantum well structure with tunable and reversible
electronic structure is formed on an individual nanotube and its
band-lineup is determined from first-principles. The elastic energy
due to the radial deformation and elastic constants are calculated and
compared with classical theories.

\end{abstract}



\maketitle

\section{Introduction}

Modification of electronic properties of condensed systems by an applied
external pressure or strain in the elastic range have been subject of
active study. However, in most of the cases, the changes one can induce
by the elastic deformation is minute even negligible due to the rigidity
of the crystals. On the other hand, the situation is rather different for
single wall carbon nanotubes (SWNTs) owing to their tubular
geometry.\cite{iijima,dressel,dressel1,mintmire1,%
hamada,mintmire2,condrev,zyao,kleiner}
SWNTs are highly flexible and have a very large Young's modulus. They
sustain remarkable elastic deformations, and it has been shown that the
structure and electronic properties undergo dramatic changes by the these
deformations.\cite{rfort1,rfort2,liu,tombler,chkov,crespi,%
sltat,ilink,cihan1,cihan2,bezryadin}
Similarly, significant radial deformation of SWNTs can be realized in the
elastic range, whereby the curvature is locally changed. This way, zones
with higher and lower curvatures relative to the undeformed SWNT can be
attained on the same circumference. Hence, one expects that radial
deformation can induce important modifications in the electronic and
conduction properties of
nanotubes.\cite{shen,lordi,feng,koredist,mazzoni,cetin,hssim}

Tight-binding calculations have indicated that a SWNT may undergo
an insulator-metal transition under a uniaxial or torsional
strain~\cite{cihan1,cihan2}. Multiprobe transport
experiments~\cite{bezryadin} on individual SWNTs showed that the
electronic structure can be modified by bending the tube, or by applying
a circumferential deformation. Empirical Extended H\"{u}ckel
calculations~\cite{rfort1} predicted that the conductance of an armchair
SWNT can be affected by the circumferential deformations and a band gap
can develop on a metallic armchair SWNT upon twisting. The effect of the
radial deformation and squeezing have been investigated by using various
methods.\cite{shen,lordi,feng,koredist,mazzoni,cetin,hssim}
However, in spite of these theoretical studies\cite{koredist,mazzoni},
a systematic analysis of the effect of the radial deformation on the
electrical properties has not been carried out yet.

The objective of this paper is to provide a better understanding of the
effect of radial deformation on the electronic band structure and elastic
properties of SWNTs,  based on the extensive first-principle
(\textit{ab-initio}) total energy and electronic structure calculations
with fully optimized structures. 
In the next section, a brief review of the the first principle
pseudopotential plane wave method that we used will be given.
The effect of the radial deformation on the atomic structure will be
discussed in Sec.~\ref{sec:geometry}. We discuss the elastic properties
of SWNTs under radial strain in Sec.~\ref{sec:elasticity}.
We show that the calculated elastic deformation energies as a function of
radial strain can be described very well within the classical theory of
elasticity. In Sec.~\ref{sec:electstrc}, we discuss the effect of the
radial deformation on the electronic properties. We find that zigzag
nanotubes are metallized under radial deformation in the elastic range.
In Sec.~\ref{sec:quantum}, this property is exploited to realize various
quantum well structures on a single nanotube with tunable electronic
properties. We applied two different radial deformations to two adjacent
regions of a $(8,0)$ nanotube to generate band offsets at the interface,
which in turn lead to multiple quantum well structures. Our conclusions
are given in Sec.~\ref{sec:conc}.

\section{Methodology}
\label{sec:method}

The first principles total energy and electronic structure calculations
have been performed using the pseudopotential plane wave
method~\cite{castep} within the generalized gradient approximation
(GGA)~\cite{gga}. Calculations have been carried
out within periodically repeating supercell geometry because of the
necessity of using the periodic boundary conditions with the plane wave
method. We used a tetragonal supercell with lattice constants, $a_{sc}$,
$b_{sc}$ and $c_{sc}$. The lattice constants, $a_{sc}$ and $b_{sc}$, are
chosen such that the interaction between nearest neighbor tubes is
negligible (the minimum C\textendash C distance between two nearest
neighbor tubes is taken as 6.2~\AA{}). The lattice constant along the axis
of the tube, $c_{sc}$, is taken to be equal to the 1D lattice parameter,
$c$, of the tube. The tube axis is taken along the
$z$\textendash direction, and the circular cross\textendash section lies
in the $(x,y)$\textendash plane. In the 1D Brillouin zone (BZ), the wave
vector $k_{z}$ varies only along the $z$\textendash axis.

Plane waves up to an energy cutoff of 500 eV are used. With this energy
cutoff and using ultra soft pseudopotentials for carbon atoms~\cite{usps},
the total energy converges within 0.5 meV/atom. In addition to this,
finite basis set corrections\cite{finbasis} are also included. Owing
to the very large lattice constants of the supercell, $a_{sc}$ and
$b_{sc}$, $\mathbf{k}$\textendash point sampling is done only along the
tube axis. The Monkhorst-Pack special $\mathbf{k}$\textendash point
scheme\cite{monpack} with with 0.02 \AA{}$^{-1}$ k-point spacing
resulting 5 and 10 {\bf k}-points within the irreducible BZ of the
tetragonal supercell are used for $(n,0)$ and $(n,n)$ tubes, respectively.
 
\section{Radially Deformed Nanotubes}
\label{sec:structure}

\subsection{Geometric Structure}
\label{sec:geometry}

\begin{figure}
\includegraphics[scale=0.33,angle=-90]{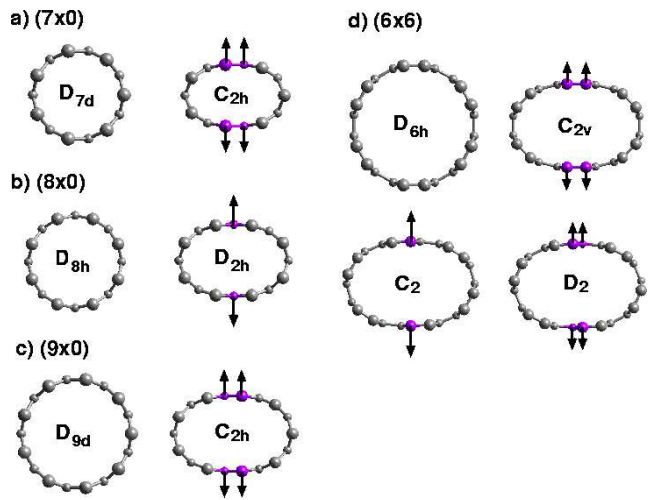}
\caption{Top view of the undeformed and deformed nanotubes.
The arrows indicate restoring forces on the fixed carbon atoms.
The in-plane mirror symmetry can be broken depending on the
rotational orientation of the tube (see (d)).
The corresponding point groups of nanotubes are also indicated.}
\label{fig:geom}
\end{figure}

The radial deformation that is treated in this study is generated
by applying uniaxial compressive stress $\sigma_{yy}$ on a narrow
strip on the surface of a SWNT. In practice such a deformation
can be realized by pressing the tube between two rigid and flat surfaces.
As a result, the radius is squeezed in the $y$\textendash direction, while
it is elongated along the $x$\textendash direction, and hence the circular
cross section is distorted to the elliptical one with major and minor axis
$a$ and $b$, respectively. A natural variable to describe the radial
deformation is the magnitude of the applied strain along the two axes
\begin{equation}
\epsilon_{yy} = \frac{R_0 - b}{R_0} ,
\end{equation}
and
\begin{equation}
\epsilon_{xx}=\frac{R_0 -a}{R_0}
\end{equation}
where $R_0$ is the radius of the undeformed (zero strain) nanotube.
We note that the point group of the undeformed nanotubes is $D_{nh}$ or
$D_{nd}$ for $n$ even or odd, respectively. Under radial deformation
described above, the point group becomes $C_{2h}$ or $D_{2h}$.
(see Fig.~\ref{fig:geom}).  However, depending on the nanotube orientation
around the tube axes the in-plane mirror symmetry can be broken. For the
$(6,6)$ tube, we studied several different orientations in order to
investigate the effect of mirror symmetry on the band crossing at the
Fermi level. Three different orientations  with point groups
$C_{2v}$, $C_2$ and $D_2$ are shown in Fig.~\ref{fig:geom}d.

For different values of strain $\epsilon_{yy}$, we carried out full
structural optimization under the constraint that the minor axis was
kept fixed at a preset value. The strains are in the elastic range,
since the deformed tubes relax back to the undeformed state when the
applied strain is removed. The structural relaxation is done in following
steps: first, depending on rotational orientation of the SWNT, either a
single bond or a carbon atom at both ends of the minor axis are pressed
towards each other by $(1-\epsilon_{yy})R_0$ and are kept fixed. Then,
under this constraint, the coordinates of the remaining atoms and the
lattice parameter of the tube $c$ are optimized. At this step, some
resultant forces remain on the fixed atom(s) with components opposite to
the applied strain as well as perpendicular to it. In the second step, the
fixed bond lengths are optimized together with all the internal coordinates
of the atoms and $c$ parameter. Eventually, in the final fully relaxed
structure the only remaining force is the restoring force, opposite to
the applied strain, on the fixed atoms. All other force components on
these fixed carbon atoms and all the forces on the rest of carbon atoms
are optimized to be less than $0.01~eV/$\AA. Figure~\ref{fig:geom} shows
the cross\textendash sectional view perpendicular to the tube axis of
the fully optimized undistorted and distorted SWNTs as well as the
restoring force vector.

\begin{figure}
\includegraphics[scale=0.52]{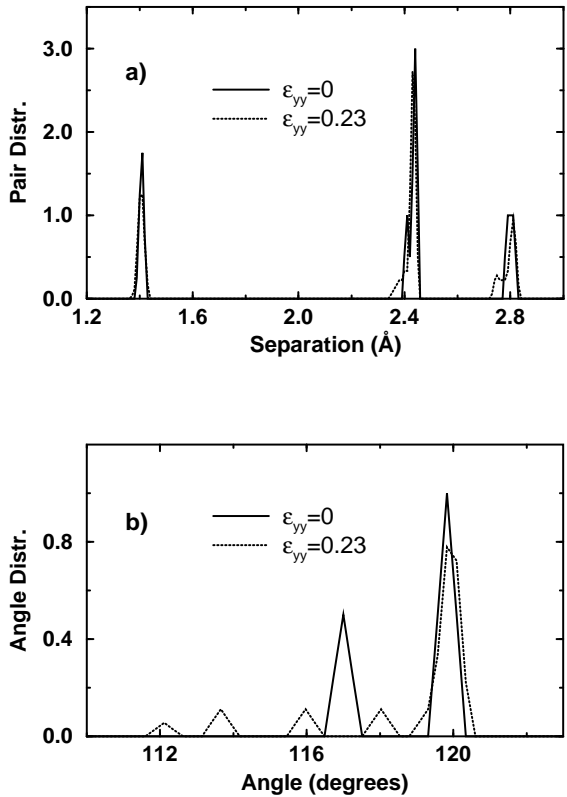}
\caption{(a) Pair distribution; (b) bond angle distribution functions
for the $(7,0)$ SWNT. Solid line is for undeformed SWNT while the dotted
line is for radially deformed one.}
\label{fig:pairdist}
\end{figure}

Figure~\ref{fig:pairdist}a shows the pair distribution function in a
deformed and undeformed (7,0) SWNT. The first peak in
Fig.~\ref{fig:pairdist}a corresponds to the first nearest neighbor
distance, which is slightly broadened without a shift of the peak position
with deformation. This indicates that the C\textendash C bond lengths
($\approx$ 1.41~\AA{}) are practically unaltered under the applied strain.
Similarly, the second peak in Fig.~\ref{fig:pairdist}a is also slightly
broadened, indicating a small effect of the distortion on the second
nearest neighbor distances. The effect of the radial deformation becomes
apparent only for the third and further nearest neigbor distances. 

Given that first and second nearest neighbor distances did not
change significantly with the radial deformation, the only remaining
degrees of freedom is the bond angle as clearly seen from the angular
distribution function shown in Fig.~\ref{fig:pairdist}b. The main peak
around $120^{\circ}$ does not change with applied strain, but the other
peak a few degrees below the main peak for undeformed tube splits into
$j$ new peaks where $j$ is the number of peaks in the radius distribution
of zigzag $(7,0)$, $(8,0)$ and $(9,0)$ SWNTs. On the other hand, for the
armchair (6,6) SWNT, although the main peak is not changed with strain,
the second peak is broadened by a few degrees. One direct consequences of
this observation is that for zigzag tubes the lattice parameter $c$
decreases very slightly with radial strain, whereas it is almost constant
for the (6,6) SWNT.

In summary, the radial deformation does not have a noticeable effect on
the first and second nearest neighbor C\textendash C distances but it
induces significant changes in the bond angles. This observation is
therefore important and has to be taken into account in tight binding
studies of SWNTs with radial deformation.

\subsection{Elasticity}
\label{sec:elasticity}

\begin{figure}
\includegraphics[scale=0.53]{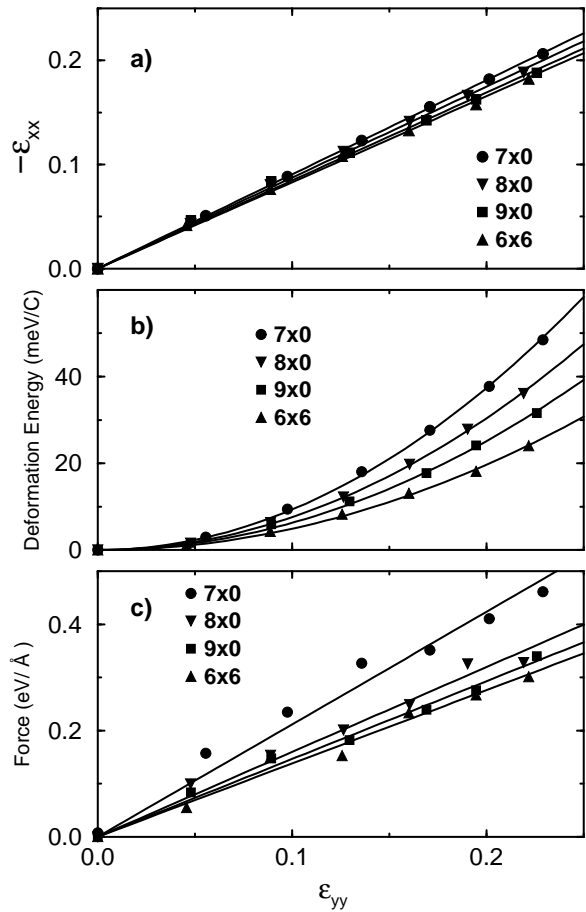}
\caption{a) The strain component $\epsilon_{xx} = (R_0 - a)/R_0$
along the major axis as a function of applied strain 
$\epsilon_{yy} = (R_0 - b)/R_0$. The slope is the in-plane Poisson
ratio, $\nu_{\parallel}$. 
(b) Variation of the elastic deformation energy per carbon atom,
(c) The restoring force on fixed carbon atoms. For $(8,0)$ SWNT,
the force is scaled by 0.5 since it is only on one carbon atom, 
while for the other tubes it is on two carbon atoms.}
\label{fig:stenfor}
\end{figure}

In order to describe the in-plane elasticity and deformation of the SWNTs,
we use first-principle calculations of the elastic deformation energy,
\textit{i.e.} the amount of energy stored in a SWNT as a result of radial
deformation, and the classical theory of elasticity. The relation between
stress and strain is given by generalized Hooke's law, for the radial
deformation described in the previous section
\begin{equation}
\sigma_{xx} = 0 = C_{11} \epsilon_{xx} + C_{12} \epsilon_{yy}
\label{eq:nu}
\end{equation}
and
\begin{equation}
\sigma_{yy} = \frac{F_y}{A} = C_{12} \epsilon_{xx} 
                            + C_{11} \epsilon_{yy}
\label{eq:stst}
\end{equation}
where $F_y$ is the restoring force applied on the surface area $A$.
$C_{11}$ and $C_{12}$ are the in-plane elastic stiffness constants.
Assuming the validity of the Hooke's law, the strain energy becomes
a quadratic function of strain as,
\begin{eqnarray}
E_{T}(\epsilon_{xx},\epsilon_{yy}) &=& E_{T}(0)  \nonumber \\
  +   \frac{1}{2} \Omega 
    (  C_{11} \epsilon^2_{xx} &+& C_{11} \epsilon^2_{yy} 
    { } +{ } 2 C_{12} \epsilon_{xx}\epsilon_{yy} ) .
\label{eq:sten}
\end{eqnarray}
The in-plane Poisson's ratio, $\nu_{\parallel}$, relates $\epsilon_{xx}$
and $\epsilon_{yy}$, from Eq.~\ref{eq:nu}
\begin{equation}
\nu_{\parallel} = -\frac{\epsilon_{xx}}{\epsilon_{yy}}
                = \frac{C_{12}}{C_{11}} .
\end{equation}
The strain components are plotted in Fig.~\ref{fig:stenfor}a. As presented
in Table~\ref{table:elastic}, $\nu_{\parallel}$ decreases with increasing
nanotube radius and is slightly smaller than 1.0.
Eqs.~\ref{eq:stst}~and~\ref{eq:sten} can be cast in a simpler form
by introducing $\nu_{\parallel}$:
\begin{equation}
\sigma_{yy} = \frac{F_y}{A} = C_{11}(1-\nu^2_{\parallel}) \epsilon_{yy} 
                            = C_{eff} \epsilon_{yy}
\label{eq:ststp}
\end{equation}
and
\begin{equation}
E_{D}= \Omega 
    \left[ \frac{1}{2}(1+\nu^2_{\parallel})C_{11} - \nu_{\parallel}C_{12}
     \right] \epsilon^2_{yy} 
\label{eq:stenp}
\end{equation}
where $E_D$ is the elastic deformation energy obtained from the difference
between the total energies of radially deformed and undeformed SWNTs
expressed in Eq.~\ref{eq:sten} (\textit{i.e}
$E_{T}(\epsilon_{xx},\epsilon_{yy}) - E_{T}(0)$).
At this point we examine how the stress and the elastic deformation
energy calculated from first principles compare with the linear and
quadratic forms in Eqs.~\ref{eq:ststp}~and~\ref{eq:stenp} obtained from
classical theory. To this end, we plot $E_{D}$ and the corresponding
restoring forces, $F_{y}$, as a function of $\epsilon_{yy}$ in
Fig.~\ref{fig:stenfor}b~and~c, respectively. Interestingly, the quadratic
form obtained from classical theory fits very well to the elastic
deformation energy calculated from the first-principles. Hooke's relation,
and hence elastic character of the deformations, persists up to
$\epsilon_{yy}=0.25$. It is also noted that the SWNT becomes stiffer as
$R$ decreases. The variation of the restoring forces is expected to be
linear in the elastic range. The restoring forces in
Fig.~\ref{fig:stenfor}c are in overall agreement with this argument,
except the deviations at certain data points due to uncertainties in
the first-principle calculations, which are amplified because the force
is a derivative quantity. Calculated elastic constants are listed in
Table~\ref{table:elastic}. It is interesting to note that there are
discrepancies in the theoretical results for Young's modulus, due to the
assignment of the \textit{thickness}, $h$, of the tube wall~\cite{portal}.
Two commonly used values are 3.4~\AA{} (based on graphite interlayer
spacing) and 0.6~\AA{} (based on the $\pi$ orbital extent). The wall
thickness $h$ can be estimated from the present radial deformation data
first by calculating the volume, $\Omega$ from Eq.~\ref{eq:stenp}.
Then, $h$ is solved by assuming that the tube is a slab with thickness $h$.
From this analysis, we found that $h$ is radius dependent and it decreases
from 0.88~\AA{} for (7,0) tube to 0.74~\AA{} for (6,6) tube.

\begin{table}
\caption{In\textendash plane elastic constants of SWNT's. All elastic
constants are in GPa except $\nu_{\parallel}$ which is unitless.
$C_{eff}=C_{11}(1-\nu^2_{\parallel})$.}
\label{table:elastic}
\begin{ruledtabular}
\begin{tabular}{l|ccccc}
 & Radius (\AA) & $\nu_{\parallel}$ & $ C_{eff}$ & $ C_{11} $ & $C_{12}$ \\
\hline
(7,0) & 2.76 & 0.904 & 129.88 & 713.36 & 645.15 \\
(8,0) & 3.14 & 0.874 & 98.70 & 416.88 & 364.20 \\
(9,0) & 3.52 & 0.864 & 91.02 & 319.67 & 270.36 \\
(6,6) & 4.06 & 0.828 & 86.12 & 273.46 & 226.34 \\
\end{tabular}
\end{ruledtabular}
\end{table}

\subsection{Electronic Structure}
\label{sec:electstrc}

\begin{figure}
\includegraphics[scale=0.53]{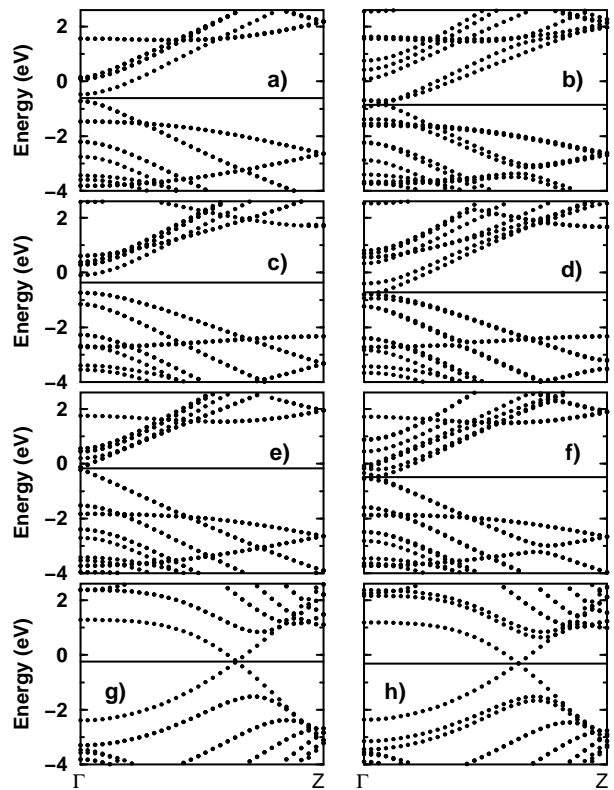}
\caption{Energy band structures of undeformed (left) and radially
deformed ($\epsilon_{yy} = 0.23$) (right) SWNTs along the
$\Gamma - Z$-direction:
a) and b) $(7,0)$,
c) and d) $(8,0)$,
e) and f) $(9,0)$,
g) and h) $(6,6)$.
Solid line is the Fermi level.}
\label{fig:band}
\end{figure}

\begin{figure}
\includegraphics[scale=0.52]{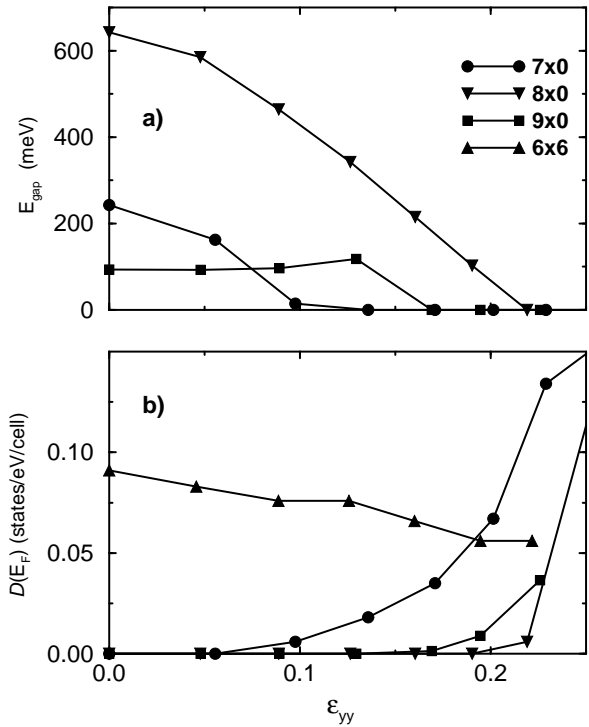}
\caption{The variation of the band gap, $E_g$ (a), and the density of
states at the Fermi level $\mathcal{D}(E_F)$ (b) as a function of applied
strain $\epsilon_{yy}$.}
\label{fig:gap}
\end{figure}

We now discuss in detail the electronic structure of SWNTs under
applied radial strain. 
The calculated band structures of undeformed and radially deformed
zigzag $(7,0)$, $(8,0)$, $(9,0)$ and armchair $(6,6)$ SWNTs are
presented near the Fermi level in Fig.~\ref{fig:band}. The band
gaps of zigzag tubes reduce with applied strain, and eventually vanish
leading to metallization. Figure~\ref{fig:gap} summarizes the variation
of band gap and density of states at the Fermi level, $\mathcal{D}(E_{F})$,
as a function of the applied strain. For $(7,0)$ and $(8,0)$ SWNT's the
band gaps decrease monotonically and the onset of an insulator-metal
transition follows with the band closures occurring at different values of
strain. Upon metallization $\mathcal{D}(E_{F})$ increases with increasing
strain. The behavior of the $(9,0)$ tube is, however, different.
Initially, the band gap increases with increasing strain, but then
decreases with strain exceeding a certain threshold value and eventually
diminishes. For all these zigzag SWNTs the band gap strongly
depend on the magnitude of the deformation, and $E_g$ is closed at
13~\%, 22~\% and 17~\% strain for (7,0), (8,0) and (9,0) nanotubes
respectively. Whereas, the armchair $(6,6)$ SWNT, which is normally
metallic, remains metallic with a slowly decreasing $\mathcal{D}(E_{F})$
even for significant radial deformation. Earlier
Delaney et al.~\cite{pseudogap1,pseudogap2} showed that the
$\pi^{*}$-conduction and $\pi$-valence bands of a $(10,10)$ tube which
normally cross at the Fermi level with quasi linear dispersion, open a
pseudogap in the range of $\sim 0.1$ eV at certain directions
of the BZ perpendicular to the axis of the tube owing to tube-tube
interactions in a rope. The opening of the gap is caused by the
broken mirror symmetry. Lammert et al.~\cite{crespi} pointed out
the gapping by squashing $(20,20)$ and $(36,0)$ metallic tubes,
since circumferential regions are brought into close proximity.
Uniaxial stress of a few kilobars can reversibly collapse a small
radius tube inducing a 0.1 eV gap, while the collapsed large radius
tubes are stable. In the study of Park \textit{et al}.~\cite{koredist},
the bandgap of the $(5,5)$ tube were monotonically increasing
probably due to bilayer interactions, since the separation of
the two nearest wall of the tube became comparable to the
interlayer distance of graphite.

\begin{figure}
\includegraphics[scale=0.38,angle=-90]{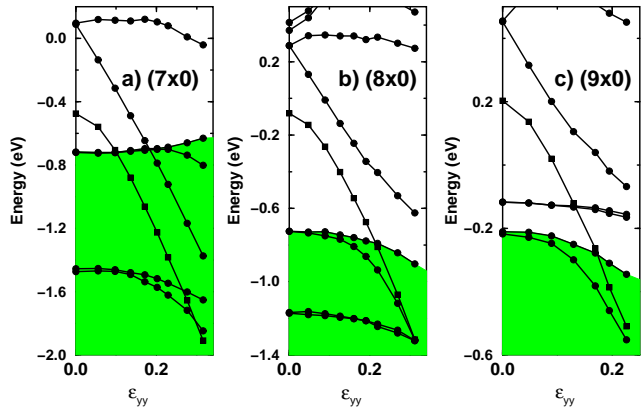}
\caption{The variation of energy eigenvalues of states near the band gap 
at the $\Gamma$\textendash point of the BZ as a function of the applied
strain. The shaded region is the valance band. The singlet state
originating in the conduction band is indicated by squares.}
\label{fig:bandab}
\end{figure}

In order to explain the band gap variation of $(n,0)$ tubes, the energies
of a few bands near the band gap are plotted as a function of strain in
Fig.~\ref{fig:bandab}. The singlet $\pi^{*}$\textendash state in the
conduction band shifts downwards in energy much faster than the other
states do with increasing strain. This is due to the increasing curvature
with increasing radial deformation. Since the singlet
$\pi^*$\textendash state lies below the double degenerate
$\pi^*$\textendash states for both $(7,0)$ and $(8,0)$ SWNT's, their
band gaps are closed monotonically with increasing $\epsilon_{yy}$.
On the other hand, for the $(9,0)$ SWNT this singlet
$\pi^*$\textendash state is above the double degenerate
$\pi^*$\textendash states. The increase of the band gap at the initial
stages of radial deformation is connected with relatively higher rate
of downward shift of the double degenerate $\pi$\textendash valence
band relative to the $\pi^{*}$\textendash conduction band under low
strains. Once the singlet $\pi^{*}$\textendash band, which shows faster
decrease with strain, crosses the doublet conduction band and enters
into the gap, the band gap begins to decrease with increasing strain.

\begin{figure}
\includegraphics[scale=0.38]{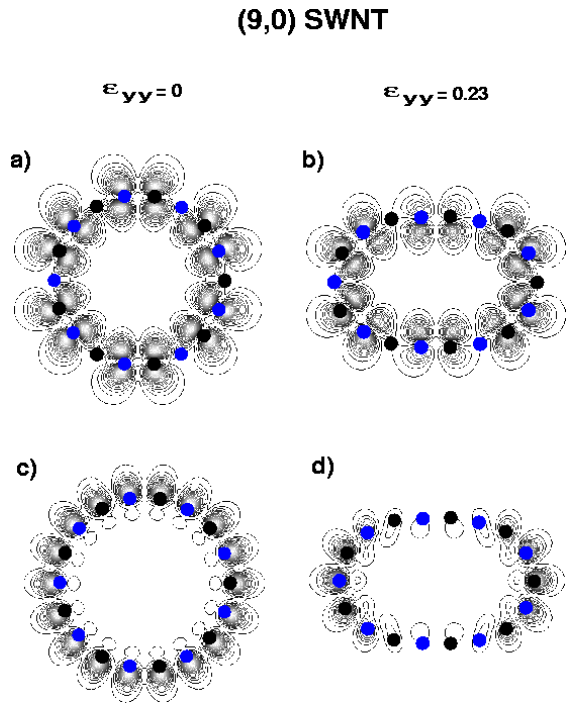}
\caption{Charge density of the highest valence band state 
at $\Gamma$\textendash point of a undeformed (a) and deformed (b)
$(9,0)$ SWNT. The panels (c) and (d) show the the singlet
``conduction''  band states.}
\label{fig:orbital}
\end{figure}

Finally, we examined the effect of the radial deformation on the
charge density. In Fig.~\ref{fig:orbital} we show the charge density of
states near the band edges. The effect of the deformation is remarkable
on the singlet state; charge moves from the low curvature regions
to the high curvature regions as the strain is increased. Significant
charge rearrangements with radial deformation can modify the chemical
activity of the surface of the SWNT relative to foreign atoms and
molecules. Since a SWNT can sustain large elastic deformations,
it allows significant charge rearrangements on its surface. Hence,
this effect can be used to control chemical reactivity of specific
carbon atoms in SWNTs~\cite{tubeadb}.

\subsection{Strain induced Quantum Structures}
\label{sec:quantum}

It is clear from the above discussion that the band gap of an
insulating SWNT can be modified, and even an insulator-metal transition
can be induced by radial deformation in the elastic range.
If the applied deformation is not uniform but has different strength
at different zones of the tube, it renders variable electronic
structure along the tube axis. For example, each zone of an individual
SWNT undergoing different radial deformation attains a different band gap.
Owing to the band off-sets at the junction, quantum structures can be
engineered on an individual tube.

Experimental and theoretical methods have been proposed in the past to
determine the band offsets, and hence to reveal the band diagram
perpetuating along the superlattice axis~\cite{esaki}. However, there are
some ambiguities in the case of SWNT superlattices $(A_nB_m)$, which are
formed by periodically repeating undeformed $A$ regions (of $n$ unit cells)
 and radially deformed $B$ regions (of $m$ unit cells):
(i) The alignment of the valence and conduction bands of the $A$ and $B$
regions and the resulting band-lineup is a complex process involving charge
transfer between $A$ and $B$, and also modification of the crystal potential
in the deformed region. In fact, the above analysis of radial
deformation has shown that the valence and conduction band edges can be
lowered when a zigzag SWNT is radially deformed. Therefore, a realistic
treatment of the band alignment requires self-consistent calculation
of the crystal potential. (ii) Even if the band diagram were known,
it is not obvious whether the Effective Mass Approximation (EMA) is
applicable for an individual, nonuniformly deformed SWNT. Therefore,
instead of applying EMA to a 1D real space band diagram or quantum well
structure, one has to perform electronic structure calculations on the
$(A_nB_m)$ supercell. However, ab-initio calculations become tedious
for large supercell size owing to many involved carbon atoms. Earlier,
we used a tight binding method and showed that states at the band edges
are confined in either $A$ or $B$ regions of a superlattices $(A_8B_8)$,
$(A_4B_{12})$ and $(A_{12}B_4)$ formed on an individual
$(7,0)$ SWNT~\cite{cetin}.

\begin{figure}
\includegraphics[scale=0.38]{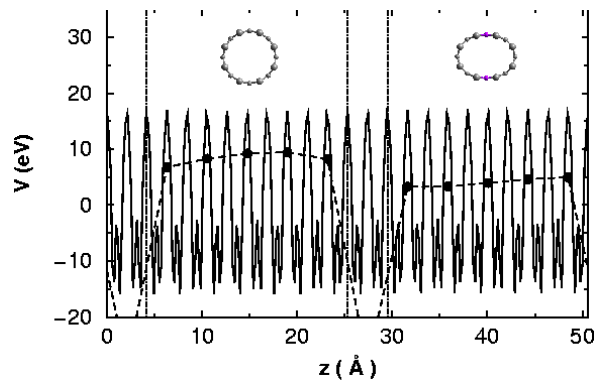}
\caption{The planar averaged crystal potential along the axis of the
$A_6B_6$ superlattice nanotube. Dotted vertical lines show the interfaces.
Circular cross\textendash section of the undeformed (8,0) nanotube in
region $A$ and elliptical cross\textendash section of radially deformed
nanotube ($\epsilon_{yy}=0.16$) in region $B$ are shown as inset.
The potential averaged over the original unit-cells of nanotube is shown
by circles and dashed line. This average potential scaled by 25
for clarity.}
\label{fig:slpot}
\end{figure}

In this study, we extend our earlier calculations of quantum structures
on SWNTs and present an ab-initio analysis of band lineup of the
$(A_6B_6)$ superlattice formed on an individual $(8,0)$ SWNT. Similar to
the previous model, here the $A$ region is left undeformed, but the $B$
region is radially deformed by $\epsilon_{yy}=0.16$. This system consists
of 384 carbon atoms in a supercell involving 12 original unit cells of
the (8,0) SWNT. We performed a partial structural optimization for the
$(A_6B_6)$ superlattice, since the full optimization is not tractable
within a reasonable computation time. We first optimized single unit
cells in the $A$ and $B$ regions corresponding to $\epsilon_{yy}=0$
and $\epsilon_{yy}=0.16$ respectively. Then, we connected $A$ and $B$
regions smoothly by one intermediate unit cell. Finally, fully
self-consistent electronic band structure calculations were carried out
on this structure. In Fig.~\ref{fig:slpot} we show the planar $(xy)$
averaged self-consistent potential,
${\bar V_{c}(z)}=\int_{S}V_{c}(\mathbf{r}) dx dy/S$.
Here $S$ is the $xy$-cross section of the supercell. The alignment of the
valence band edges between undeformed $A$ and deformed $B$ are revealed by
first integrating the planarly averaged potential
${\bar V_{X}}=\int_l {\bar V_c(z)}dz/l$ at each region, ($X=A$ or $B$),
over a length of original unit cell (i.e. $l=c$) along the tube
axis~\cite{baroni}. In Fig.~\ref{fig:slpot}, ${\bar V_X}$ is shown for
both regions of the supercell which constitutes the reference level for
band-lineup. In the next step, we determine the energy of the valence
band edge from the ${\bar V_A^{\infty}}$ and ${\bar V_B^{\infty}}$
calculated for two different, uniform (infinite) (8,0) SWNTs (one
undeformed, the other uniformly deformed with $\epsilon_{yy}=0.16$).
These are $E_{V,A}^{\infty}$ and $E_{V,B}^{\infty}$. It is assumed that
in the nanotube superlattice $E_{V,A}^{\infty}$ and $E_{V,B}^{\infty}$
are unaltered. The band-lineup of the valence band is calculated from the
difference
$\Delta E_V={\bar V_A}-{\bar V_B}+E_{V,A}^{\infty}-E_{V,B}^{\infty}$.
For $(8,0)$ SWNT, we find $\Delta E_V\sim180$ meV; the valence band edge
of $B$ is lower than that of $A$ indicating a staggered band lineup.
This result clearly shows that by applying periodic radial deformation
on an individual semiconducting SWNT one can generate a quantum structure,
where the band gap in the direct space along the tube axis undergoes
a periodic variation which is continuously tunable and reversible.

\section{Conclusions}
\label{sec:conc}

In this work we present an extensive first principle analysis of the
effect of radial deformation on the atomic structure, energetics and
electronic structure of SWNTs. We find that the energy band
structure and the variation of the gap with radius (or $n$) differs
from what one derived from the zone folded band structure of graphene
based on simple tight binding calculations. More interestingly,
the response of the energy bands around the band gap to the applied
radial deformation is different for different bands. Depending on the
relative position of these bands, the band gap displays different
behavior under the radial deformation. In general, the band gap is
reduced and eventually closed to yield an insulator-metal transition
under the elastic radial deformation. The strong dependence of the band
gap on the applied strain, its reversible and continuously tunable
behavior are exploited to form quantum well structure on an individual
SWNT. A first-principle calculation of the alignment of the valence
band is presented. The deformation energy and elastic constants under
the radial deformation are calculated. We find that the strain energy
due to the radial deformation can be fitted very well to the quadratic
expressions obtained from the classical theory of elasticity
within the Hooke's law.

\begin{acknowledgments}
This work was partially supported by the National Science Foundation
under Grant No. INT01-15021 and T\"{U}B\'{I}TAK under 
Grant No. TBAG-U/13(101T010).
\end{acknowledgments}

\end{document}